# Two computational regimes of a single-compartment neuron separated by a planar boundary in conductance space

Brian Nils Lundstrom[1], Sungho Hong[1], Matthew H Higgs[1,2] and Adrienne L Fairhall[1]
Department of Physiology and Biophysics, University of Washington, Seattle[1]
Veterans Affairs Puget Sound Health Care System, Seattle, Washington[2]

**Abstract:**
Recent *in vitro* data show that neurons respond to input variance with varying sensitivities. Here, we demonstrate that Hodgkin-Huxley (HH) neurons can operate in two computational regimes, one that is more sensitive to input variance (differentiating) and one that is less sensitive (integrating). A boundary plane in the 3D conductance space separates these two regimes. For a reduced HH model, this plane can be derived analytically from the *V* nullcline, thus suggesting a means of relating biophysical parameters to neural computation by analyzing the neuron's dynamical system.

**Introduction**

A neuron's ion channel configuration determines how it processes information. However, the relationship between a particular set of ion channels and the specific computation it performs is still not clear, even for basic Hodgkin-Huxley (HH) model neurons (Hodgkin and Huxley, 1952; Aguera y Arcas et al., 2003; Yu and Lee, 2003) and simplified versions of the model (Kepler et al., 1992; Gerstner and Kistler, 2002; Murray, 2002). In this paper, we examine information processing in the HH neuron, where this model neuron is taken to be a generic case with typical spike-triggering ion channels. We focus on the response of the space-clamped HH model to a time-varying synaptic current input.

We approximate the net synaptic current input at the soma as exponentially-filtered Gaussian white noise current (Gerstein and Mandelbrot, 1964; Bryant and Segundo, 1976; Tuckwell, 1988; Mainen and Sejnowski, 1995; Destexhe et al., 2001; Rauch et al., 2003; Rudolph and Destexhe, 2003a; Richardson and Gerstner, 2005), in which the input mean reflects the average number of inputs while fluctuations, quantified by the variance, generally depend on the degree of neuronal input synchrony (Destexhe and Pare, 1999; Moreno et al., 2002; Richardson, 2004). We use the firing rate-current ($f$-$I$) function as a straightforward means of examining the model's input/output response properties. If firing rate varies most strongly with mean current, we term the neuron an *integrator*; if the firing rate is sensitive to variance and relatively insensitive to the mean current, the neuron is fluctuation-driven, and we refer to it as a *differentiator* (Abeles, 1982; Konig et al., 1996; Higgs et al., 2006). A differentiating neuron is thus characterized by low or zero firing rate in response to a constant or zero variance input.

Naturally, a spectrum exists between these two classifications. In addition, the operating mode of a given neuron often depends on the mean input current. Many neurons function as differentiators at low, subthreshold mean currents but behave more like integrators at suprathreshold mean currents; other neurons differentiate their input at saturating mean currents. However, the purest neural differentiators, such as specialized coincidence detectors in the auditory brainstem (Oertel, 1983; Reyes et al., 1994; Rothman and Manis, 2003) appear entirely incapable of integration, and never fire repetitively at any level of constant current stimulation.

Modeling studies suggest that neurons can function as either integrators or coincidence detectors (differentiators) based on the nature of incoming input (Gutkin et al., 2003; Rudolph and Destexhe, 2003b). *In vitro* studies suggest that high conductance inputs facilitate coincidence detection, or differentiation (Gonsalves and Paller, 2000; Destexhe et al., 2003; Prescott et al., 2006), and at least one *in vivo* study suggests that synchronized inputs more effectively drive cortical spiking (Roy and Alloway, 2001). Recent *in vitro* studies suggest that neuronal firing rates are affected by the statistical properties of approximated synaptic inputs (Chance et al., 2002; Fellous et al., 2003), and that different populations of neurons respond to input variance with differing sensitivities (Higgs et al., 2006; Arsiero et al., 2007).

Our aim was to determine the biophysical parameter regime for which the HH neuron retains sensitivity to input variance even at high input means. We conclude that the HH neuron can process information in two fundamental ways, as both an integrator and a differentiator, and that, despite the highly nonlinear nature of the model, a planar boundary in the space of maximal conductances separates these two regimes. Further, using a 2D simplification of the HH model, we find that this planar boundary can be derived directly from model equations and related dynamical system properties, thus demonstrating a simple link between the observed categories of computational function and the biophysical conductance parameters.

**Hodgkin-Huxley neuron as both integrator and differentiator**

Recent studies show that *f-I* curves from *in vitro* cortical pyramidal cells, hippocampal CAI pyramidal cells, and neurons from avian auditory brainstem demonstrate changes in sensitivity to input variance as the mean is increased, and suggest that these changes are related to slow adaptation currents (Higgs et al., 2006; Prescott et al., 2006) or slow sodium inactivation (Arsiero et al., 2007). However, many simple neuron models, including the standard Hodgkin-Huxley (HH) model neuron, the Connor-Stevens model neuron, and the Traub-Miles model neuron, show decreased sensitivity to input variance as the mean increases. We have found that for several simple model neurons, lowering the maximal sodium conductance $G_{Na}$, such as through voltage-dependent slow sodium inactivation, leads to an increased sensitivity to input variance at high means, as would be the case for differentiating neurons (Figure 1). In this differentiator operating regime, these neurons do not fire repetitively to constant, or zero-variance, inputs.

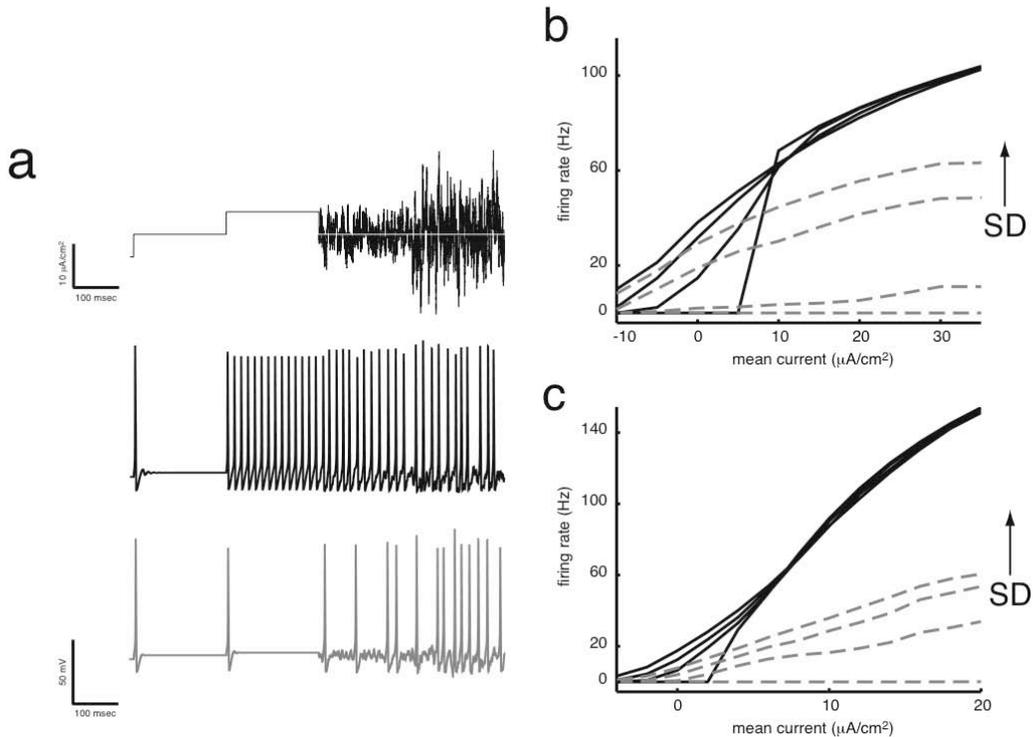

**Figure 1: Spectrum of integration and differentiation. (a) Simple differentiators do not respond to zero variance inputs at steady state.** This non-zero current stimulus undergoes a two-fold increase in mean, [5 10] µA/cm², followed by a two-fold increase in standard deviation, [2.5 5] µA/cm², and is presented to two different neuron models. Although not ideal, the black trace has integrator characteristics, whereby changes in standard deviation result only in very small firing rate changes. In contrast, firing rates in the gray trace increase as standard deviation increases, but the steady state firing rate is always zero if input variance is zero. **Decreased sodium conductance leads to neurons that are better differentiators in (b) the standard Hodgkin-Huxley model neuron and in (c) a neocortical model neuron (p. 124, Gerstner and Kistler, 2002).** Data are plotted for the Hodgkin-Huxley and neocortical models with high (black, solid lines) and low (gray, dotted lines) sodium conductances. Lines represent mean firing rate at steady state in response to input current; increased input standard deviation generally leads to increased firing rate. Models with low sodium conductance do not fire in response to zero-variance inputs at steady state. Parameters for (b) HH: $G_{Na}$ = [120 82] mS/cm², SD = [0 2 4 6]

μA/cm² and (c) cortical: $G_{Na}$ = [50 25] mS/cm², SD = [0 2 3 4] μA/cm² neurons. Standard parameters (Appendix A) were used unless otherwise noted.

Here, we consider only static maximal conductance values rather than any activity-dependent (LeMasson et al., 1993; Giugliano et al., 1999) or voltage-dependent change. The membrane potential (*V*) of the space-clamped Hodgkin-Huxley (HH) neuron is described by

$$C\frac{dV}{dt} = -G_{Na}m^3h(V - E_{Na}) - G_K n^4(V - E_K) - G_{Leak}(V - E_{Leak}) + I,  \quad (1)$$

where *C* is capacitance, $G_i$ are the maximal channel conductances, $E_i$ are the reversal potentials, *I* is the external, injected current, and *[m, h, n]* are the channel gating variables, which obey first order dynamics (Appendix A). If we take this standard HH neuron and systematically lower its maximal sodium conductance, its response to inputs, as evaluated by an *f-I* curve, changes (Figure 2). When $G_{Na}$ is high, the HH neuron behaves more like an integrator, while when $G_{Na}$ is low, the neuron behaves more like a differentiator.

Conceptually, these differences in the *f-I* curves (Figure 1, gray and black) result from changes in an effective voltage threshold for spiking, where spike initiation occurs when enough sodium channels open to counter outward currents. The maximal conductance $G_{Na}$, for example, approximately establishes the spiking threshold. Additionally, the availability of sodium channels depends on the voltage-dependent inactivation variable *h*, adjusting the effective threshold value. Differentiation then occurs when slow components of the input are below this threshold, and firing only results from input noise. In certain regimes, such as when $G_{Na}$ has been lowered sufficiently, this effective threshold is never surpassed as input mean increases – the voltage-dependent threshold keeps increasing slightly as the input mean increases until eventually the neuron is unable to spike. This effect, known as depolarization block, is observed for large inputs in the *f-I* curves of Figure 2. This conceptual explanation will be further developed in terms of the neuron's dynamical system.

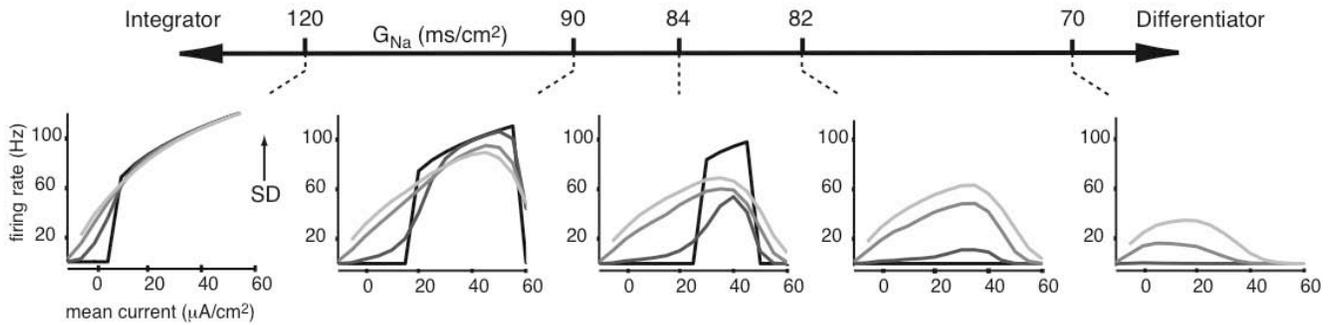

**Figure 2: The HH neuron behaves more like a differentiator as its sodium conductance $G_{Na}$ is lowered.** As $G_{Na}$ is lowered below ~83 mS/cm², the HH neuron ceases to respond to zero-variance input. Traces on each *f-I* plot from black to light gray represent increasing SD, where SD = [0 2 4 6] μA/cm². The first and fourth *f-I* plots from the left show the same data as in Figure 1b. Standard HH values ($G_{Na}$, $G_K$, $G_{Leak}$) = (120, 36, 0.3) mS/cm² were used unless otherwise noted (Appendix A).

## A planar boundary separates integration from differentiation

We simulated the HH model using a set of 56 different conductance values ($G_{Na}$, $G_K$, $G_{Leak}$) and from these determined *f-I* curves from each condition. As seen in Figure 2, at high $G_{Na}$ the HH neuron responds well to noiseless inputs, whereas at low $G_{Na}$ the neuron does not respond to zero-variance inputs. This transition from noiseless firing to no response is relatively sudden along the range of $G_{Na}$ in which firing occurs. This transition can also be observed when $G_K$ or $G_{Leak}$ are increased, or when these three conductances are modulated in linear combination according to[1]:

$$G_{Na} - 2.07 G_K - 22.8 G_{Leak} = 0, \qquad (2)$$

where conductances have units mS/cm². When this sum is less than zero, the HH neuron will not fire to a noiseless current, whereas if the sum is greater than zero, the neuron fires repetitively. Eq. (2) describes the empirically observed plane that passes through the origin in the three-dimensional space of HH conductances, which is shown as a line for fixed $G_{Leak}$ in Figure 3a. Here, if the neuron does not fire to a noiseless current, no matter how large the mean, then the neuron is considered to behave more like a differentiator[2]. In terms of the dynamical system, when the right hand side of Eq. (2) is greater than zero, the system undergoes a Hopf bifurcation at a particular mean current (e.g. approx 6.5 µA/cm² in Figure 1b, bottom black trace); however, if the right hand side is less than zero, the one fixed point of the dynamical system is stable for all mean current values (as in Figure 5b and c).

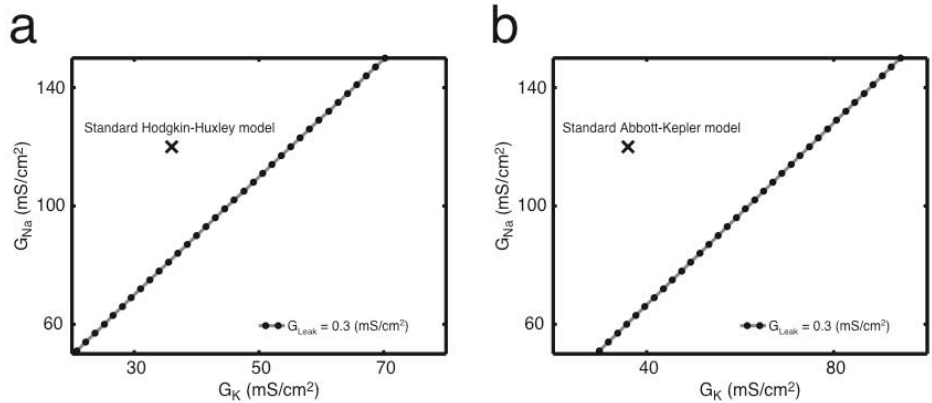

**Figure 3: The boundary between no response to constant stimuli (differentiation) and firing to constant stimuli (integration) is described by a line in the two-dimensional space of $G_{Na}$ and $G_K$.** This boundary is similar for both the Hodgkin-Huxley (**a**) and Abbott-Kepler (**b**) neuron models. As $G_{Leak}$ increases, the boundary remains linear with the same slope but an increasing intercept in $G_{Na}$. The black *X*'s designate the standard parameters for the Hodgkin-Huxley and Abbott-Kepler models.

---

[1] These coefficients are the result of a multiple least squares regression of 56 conductance sets ($G_{Na}$, $G_K$, $G_{Leak}$) determined via simulation. For a given $G_{Na}$ and $G_{Leak}$, $G_K$ would initially take a relatively low value for which the neuron would fire to some mean input current *I*. $G_K$ was then increased incrementally and *f-I* curves computed until the neuron would not fire to any mean input *I*; this was then one of the boundary sets. As the ratios of the conductances (i.e. $N = G_{Na}/G_{Leak}$ and $K = G_K/G_{Leak}$) increased, the steady state voltages $V_0$ at the boundary current value (which we will later define as *I\**) generally decreased: mean –52.8; std 2.0; range [-48.6 –54.8]. Conductances were chosen according to the following constraints: *N* was within the range [50 500], $G_{Na} > 50$ *mS/cm²*, $G_{Leak} = 0.3, 1,$ or *2*. The best fit of these data passes through or very close to the origin (Appendix B).

[2] Although for the HH model this is an appropriate criterion, other models can show non-zero firing rates in response to zero-variance stimuli while their *f-I* curves are differentiator-like as in Figure 2. Further, even neurons that are here classified as integrators, as seen on the left side of Figure 2, display differentiator-like characteristics when mean input currents are low.

Having obtained Eq. (2) through simulation, we would like to derive it directly from model equations. However, due to its four differential equations, the dynamical system of the HH model is difficult to analyze. Therefore, we use the two-dimensional Abbott-Kepler (AK) model, which is derived from the HH model equations and maintains a direct correspondence of parameters (Abbott and Kepler, 1990; Kepler et al., 1992). In the AK model, the four variables of the HH model ($V$, $m$, $h$, and $n$) are divided by time scale into two groups that become $V$ and $U$, a fast activation and a slow recovery variable, respectively. The 2D AK model equations have the form (Abbott and Kepler, 1990; Kepler et al., 1992; Hong et al., In press):

$$C\frac{dV}{dt} = -G_{Na}m_\infty^3(V)h_\infty(U)(V-E_{Na}) - G_K n_\infty^4(U)(V-E_K) - G_{Leak}(V-E_{Leak}) + I, \quad (3)$$

$$\frac{dU}{dt} = \frac{G_{Na}(V-E_{Na})m^3(V)(h(V)-h(U))/\tau_h(V) + 4G_K(V-E_K)n^3(U)(n(V)-n(U))/\tau_n(V)}{G_{Na}(V-E_{Na})m^3(V)h'(U) + 4G_K(V-E_K)n^3(U)n'(U)}. \quad (4)$$

Eq. (3) has a similar form to Eq. (1), where the dynamics of the gating variables have been replaced by their steady state voltage-dependent values $m_\infty$, $n_\infty$ and $h_\infty$. Like the HH model, this model shows a boundary plane between integration and differentiation, which is described by[3]:

$$G_{Na} - 1.54 G_K - 16.7 G_{Leak} = 0, \quad (5)$$

where conductances have units mS/cm$^2$. The line that is described by Eq. (5) when $G_{Leak} = 0.3$ mS/cm$^2$ is shown in Figure 3b.

**Deriving the planar boundary**

The advantage of working with a two-dimensional dynamical system is that we can more easily understand how the empirical boundary equation, Eq. (5), emerges from the dynamical system, Eqs. (3) and (4). We begin from the phase portrait of the AK model. The flow is determined by the $V$ and $U$ nullclines, defined as the set of points for which $dV/dt = 0$ and $dU/dt = 0$, respectively. The unique fixed point of the system is the point of intersection of the two nullclines (Gerstner and Kistler, 2002; Murray, 2002; Izhikevich, 2007). In the AK model, from Eq. (4), the $U$ nullcline is given by $U=V$. The $V$ nullcline generally takes an N-shape for intrinsically spiking neural models (Izhikevich, 2007), and is given by the solution of

$$0 = -G_{Na}m_\infty^3(V)h_\infty(U)(V-E_{Na}) - G_K n_\infty^4(U)(V-E_K) - G_{Leak}(V-E_{Leak}) + I, \quad (6)$$

which for simplicity we express as

$$f(U,V) = I, \quad (7)$$

where

$$f(U,V) \equiv G_{Na}m_\infty^3(V)h_\infty(U)(V-E_{Na}) + G_K n_\infty^4(U)(V-E_K) + G_{Leak}(V-E_{Leak}).$$

Thus, for the AK model, the shape of the $V$ nullcline in $(V,U)$ space depends nonlinearly on mean current $I$ as well as on the parameters of Eq. (7). We will focus on the three maximal conductances ($G_{Na}$, $G_K$, $G_{Leak}$).

---

[3] This process was the same as for the HH model for 55 conductance sets. Again, as conductance ratios (i.e. $N = G_{Na}/G_{Leak}$ and $K = G_K/G_{Leak}$) increased, steady states voltages $V_0$ at the boundary current value (which we will later define as $I^*$) decreased: mean –50.3; std 0.8; range [-48.2 –51.2]. The best fit of these data passes through or very close to the origin (Appendix B).

For any given set of the four parameters $(G_{Na}, G_K, G_{Leak}, I)$, whether or not the neuron fires in response to DC current is determined by two things: the shape of the *V* nullcline and the position of the fixed point on this nullcline. If the *V* nullcline has an *N*-like shape, as seen in Figure 4a and b, the system is excitable. However, for some values of *I*, the *V* nullcline can flatten and the system becomes unable to create a spike; this corresponds to depolarization block and is seen in the top contour of Figure 5a, which is the *V* nullcline for $I=40$ µA/cm² and the given conductances. When the system is able to spike, the second factor determining DC response is where the fixed point, the intersection of the *V* nullcline with the line $U = V$, lies with respect to the *local minimum* of the V nullcline or the left "knee" of the curve. When the fixed point is to the right of the knee, the neuron fires repetitively to zero-variance mean current (Figure 4a, 4c), but when the fixed point is to the left of the local minimum it does not (Figure 4b, 4d)[4].

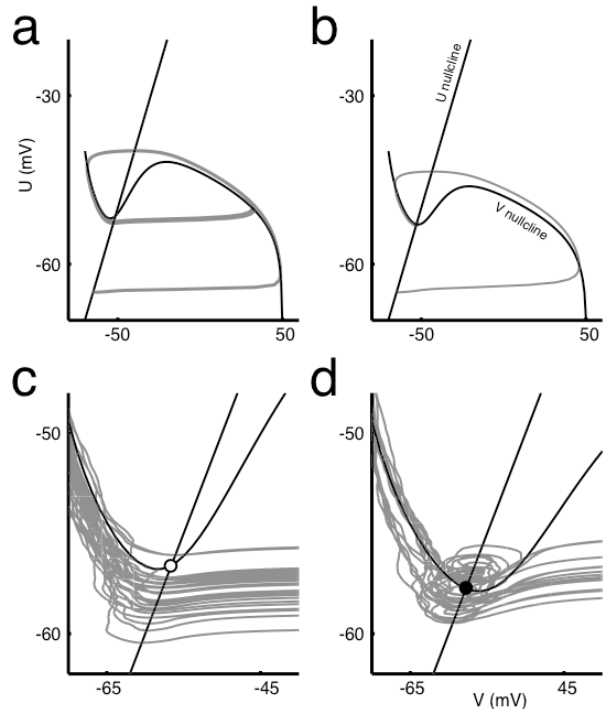

**Figure 4: Phase portraits of the Abbott-Kepler neuron model.** (**a**) The neuron fires repetitively to a mean current of 50 µA/cm² when sodium conductance is high ($G_{Na} = 120$ mS/cm²) such that the conductance values lie above the plane of Eq. (5) but (**b**) does not fire repetitively when the sodium conductance is low ($G_{Na} = 50$ mS/cm²). The straight black line is the *U* nullcline corresponding to $dU/dt = 0$, where *U* is a general recovery variable, while the curved black line is the *V* nullcline, satisfying $dV/dt = 0$, where *V* is the membrane voltage. The gray lines are the trajectories of *(V,U)* in time, where the loops in (**a**) are the limit cycles of spikes. For a mean of 20 µA/cm² and SD of 4 µA/cm², spike initiation is dominated by (**c**) the internal dynamics when $G_{Na}$ is high but (**d**) when $G_{Na}$ is low, spikes are initiated by input variance. The fixed points in the two cases are (**c**) unstable, open circle, and (**d**) stable, closed circle, respectively. For (**c**) and (**d**), $I = 20$ µA/cm². Other model parameters include: $G_K = 36$ mS/cm² and $G_{Leak} = 0.3$ mS/cm².

---

[4] This may not be strictly true. As has been noted for the Fitzhugh-Nagumo model (Izhikevich, 2007), loss of stability by the fixed point may not occur precisely at the local minimum in the *V*-nullcline, as can be verified by linearization and local stability analysis. The effect is small in the AK model (Appendix B).

When the current is noisy, the position of the fixed point with respect to the minimum determines whether spike initiation is largely dictated either by the internal dynamics of the system (Figure 4c) or by momentary, large current fluctuations (Figure 4d). These fluctuations cause leftward or rightward excursions along the V-axis, which may cause the neuron to cross an upward curving threshold function situated slightly beneath, but approximately parallel to, the V nullcline (Hong et al., In press). If a change in stimulus is gradual, the nullclines alter their geometry, which may prevent threshold crossing despite an equally large or larger stimulus perturbation.

Each point in the parameter space $(G_{Na}, G_K, G_{Leak}, I)$ maps onto one of three possible conditions: no minimum exists in the V nullcline and thus the system is not excitable; or the system is excitable, with a fixed point either to the right or the left of the minimum. For the AK model to implement a perfect differentiator, it should not fire spontaneously for any mean current level $I$. For this to be the case, for values of $I$ such that the V nullcline shows a minimum, the nullclines must have the configuration of Figure 4b or 4d. Thus, as $I$ increases, the V-nullcline minimum will disappear before the fixed point shifts to the right of the V nullcline minimum. We use this to derive the boundary condition on conductance parameters.

Let us consider an example of the behavior of the system. In Figure 5a, we show contours of the surface Eq. (7), which are a set of V nullclines each for different $I$. As $I$ increases, the nullclines lose their $N$-like shape, and eventually the neuron is no longer excitable and undergoes depolarization block. For differentiating neurons, the line $U = V$ always intersects the V nullclines at a place where the V nullcline is decreasing, i.e. the derivative of the V nullcline evaluated at the fixed point is negative: as can be seen in Figure 5, this means that $\partial f / \partial V > 0$. A boundary conductance ($G_{Na}$, $G_K$, $G_{Leak}$) gives, then, a set of $I$-dependent V nullclines for which this partial derivative has a minimum value of zero, as in Figure 5d.

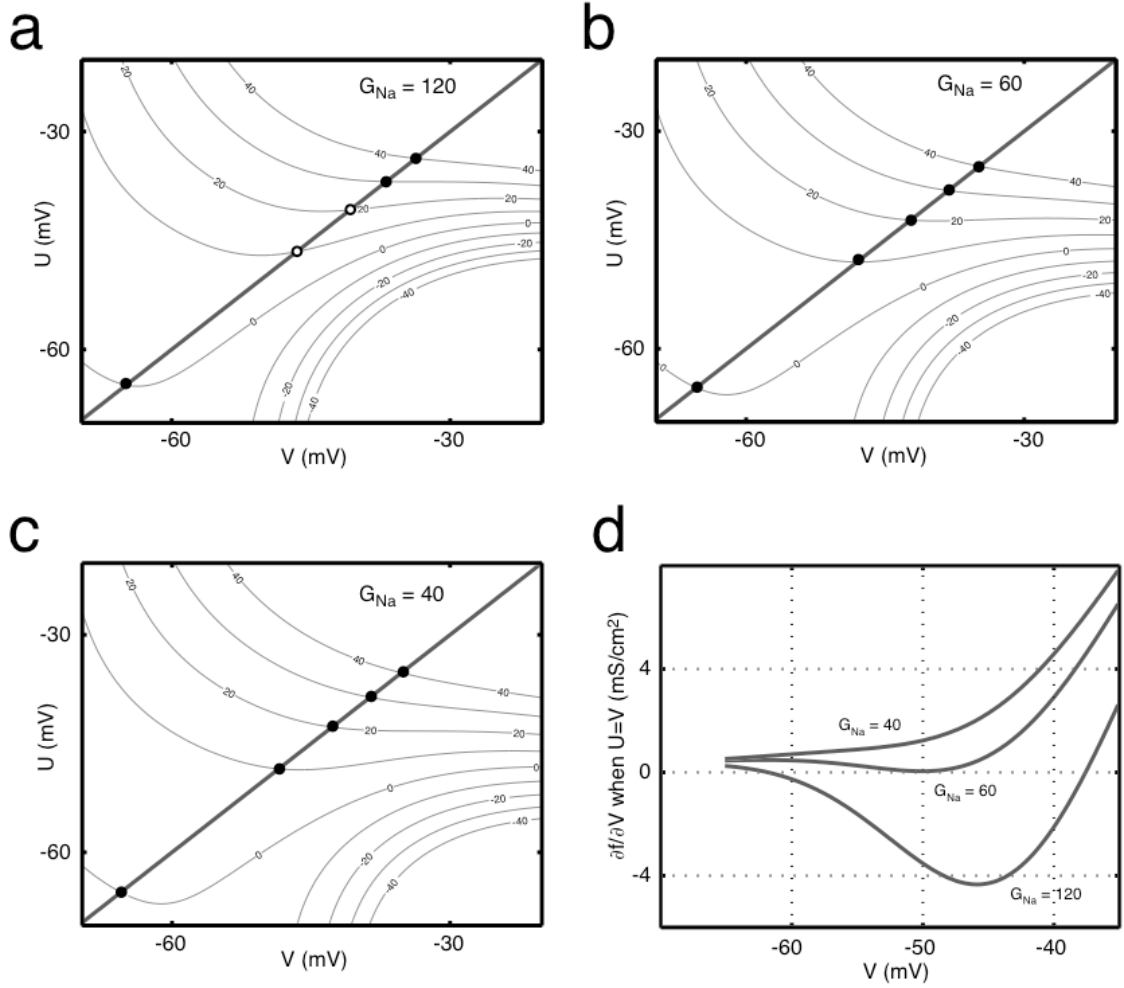

**Figure 5: $V$ nullclines described by $f(U,V) = I$. (a)** The surface of Eq. (7) is plotted as a contour in current $I$ ($\mu$A/cm$^2$) for the conductances $(G_{Na}, G_K, G_{Leak}) = (120, 36, 0.3)$. **(b)** $f(U,V)$ for $(G_{Na}, G_K, G_{Leak}) = (60, 36, 0.3)$, which is very close to the boundary. **(c)** $f(U,V)$ for $(G_{Na}, G_K, G_{Leak}) = (40, 36, 0.3)$. Fixed points fall on the line $U=V$, where stable and unstable fixed points are represented by closed and open circles, respectively. **(d) The partial derivative of $f(U,V)$ with respect to $V$ evaluated for $U=V$ for three sets of conductances.** The top ($G_{Na} = 120$ mS/cm$^2$) and bottom ($G_{Na} = 40$ mS/cm$^2$) traces represent approximately integrating and approximately differentiating neurons, respectively. The middle ($G_{Na} = 60$ mS/cm$^2$) trace lies approximately at the boundary between the two regimes. For this AK model, $(G_K, G_{Leak})$ were (36, 0.3) mS/cm$^2$.

For the fixed point to occur to the left of the minimum, we require that $dU/dV < 0$ for all fixed points $(U_0(I), V_0(I))$. Thus, the condition is given by

$$\left.\frac{dU}{dV}\right|_{U_0,V_0} = -\left.\frac{\partial f / \partial V}{\partial f / \partial U}\right|_{U_0,V_0} = 0. \tag{8}$$

One of the two fixed-point conditions is that all fixed points, independent of $I$, must satisfy $U = V$. Thus, fixed points for all $I$ fall somewhere along the line $U = V$. Since we want to characterize the system's behavior for all $I$, we will thus evaluate the condition at $U = V$. Further, we will only consider the numerator of Eq. (8) (see also Appendix B), giving the condition

$$\left.\frac{\partial f(U,V)}{\partial V}\right|_{U=V} = 0. \tag{9}$$

Figure 5d shows some examples of the function $\partial f / \partial V$, evaluated for several values of the conductances. Our boundary criteria require not only that the derivative is zero, but that zero is the minimal value for any $I$. Thus, at the point $(V_0(I^*), U_0(I^*))$, where the derivative takes the value zero, the function $\partial f / \partial V$ must be at a minimum. Hence, we require that for some value of $V$, $V^* = V_0(I^*)$, where $V^*$ is a function of $I$ and the conductance parameters, Eq. (9) and the following hold simultaneously:

$$\frac{d}{dV}\left(\left.\frac{\partial f(U,V)}{\partial V}\right|_{U=V}\right) = 0. \tag{10}$$

Because both boundary equations Eqs. (9) and Eq. (10) are homogenous in $G_{Na}$, $G_K$ and $G_{Leak}$, we can describe the boundary plane with two variables. For example, we can rewrite Eq. (9) as:

$$A_1(V^*)G_{Na} + B_1(V^*)G_K + G_{Leak} = 0$$
$$A_1(V^*)\frac{G_{Na}}{G_{Leak}} + B_1(V^*)\frac{G_K}{G_{Leak}} = -1 \tag{11}$$
$$A_1(V^*)N + B_1(V^*)K = -1$$

where $N$ and $K$ are conductance ratios. From Eq. (10), we obtain another constraining equation and must then solve the following system:

$$\begin{pmatrix} A_1(V^*) & B_1(V^*) \\ A_2(V^*) & B_2(V^*) \end{pmatrix} \begin{pmatrix} N \\ K \end{pmatrix} = \begin{pmatrix} -1 \\ 0 \end{pmatrix}, \tag{12}$$

where $A_1$ and $B_1$ are coefficients of Eq. (9) as written explicitly in Eq. (11),

$$A_1(V) = \left[m_\infty^3(V)\right]' h_\infty(V)(V - E_{Na}) + m_\infty^3(V) h_\infty(V),$$
$$B_1(V) = n_\infty^4(V),$$

and $A_2$ and $B_2$ are coefficients of Eq. (10),

$$A_2(V) = \left[m_\infty^3(V)\right]'' h_\infty(V)(V - E_{Na}) + \left[m_\infty^3(V)\right]' \left[h_\infty(V)\right]'(V - E_{Na})$$
$$+ 2\left[m_\infty^3(V)\right]' h_\infty(V) + m_\infty^3(V)\left[h_\infty(V)\right]',$$
$$B_2(V) = \left[n_\infty^4(V)\right]'.$$

The general solution is:

$$N = \frac{-B_2(V^*)}{A_1(V^*)B_2(V^*) - A_2(V^*)B_1(V^*)}$$
$$K = \frac{A_2(V^*)}{A_1(V^*)B_2(V^*) - A_2(V^*)B_1(V^*)}. \tag{13}$$

The forms of $N$ and $K$ are highly nonlinear due to the dependence on the unknown value $V^*$ (Figure 6a), where $V^* > -51$ mV for positive conductance ratios. Rather than solve for $V^*$ for every parameter set, we observe that the relevant range of $V^*$ is highly constrained by the need to generate conductance values in a physiological range. The range used, $V^* = [-50.5\ -48]$, gives $N = [50\ 500]$, which was a constraint for the simulation. In this very narrow range of $V^*$, $N$ and $K$ are linearly related to one another; these are plotted against one another in Figure 6b. When we fit the line of $N$ vs. $K$, we obtain coefficients of ~1.55 and 16.5, which are very close to the coefficients of $G_K$ and $G_{Leak}$ in Eq. (5), respectively. Exact agreement between simulated and calculated results is probably prevented by imprecision in gathering the simulated boundary points.

The coefficients of $G_K$ and $G_{Leak}$ in Eq. (5) can be determined analytically from the coefficients $A_1$ and $B_1$ of Eq. (11). Specifically, for a given value $V^*$, the coefficient for $G_K$ is $-B_1(V^*)/A_1(V^*)$ and the coefficient for $G_{Leak}$ is $-1/A_1(V^*)$, as plotted in Figure 6c and 6d. Given Eq. (13), $-B_1/A_1$ and $-1/A_1$ are the slope and y-intercept of the linear relationship $N$ vs. $K$. Regression, as in Figure 6b, takes into account many values of $V^*$ and does not require knowing $V^*$ a priori, while using $A_1$ and $B_1$ gives an analytic result for a given $V^*$.

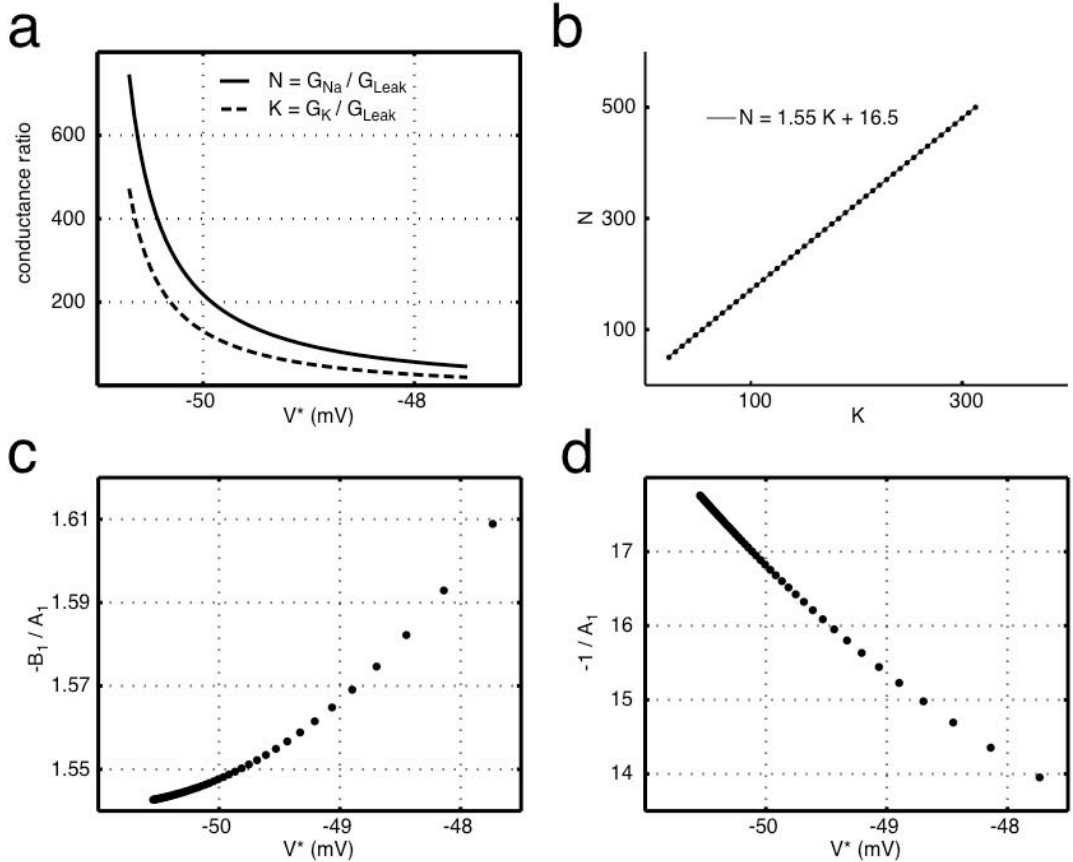

**Figure 6: Conductance ratios $N = G_{Na} / G_{Leak}$ and $K = G_K / G_{Leak}$ plotted (a) as functions of V and (b) against each other.** The coefficients of the linear regression fit are close to the simulated coefficients of Eq. (5). The voltage range plotted in (**b**) is from -50.5 to –48 mV, where lower voltages correspond to higher ratios. **Coefficients (c) $-B_1/A_1$ and (d) $-1/A_1$ of Eq. (9) can be plotted as a function of $V^*$, where these values of $V^*$ give ratios of $N = [50\ 500]$.** For a given $V^*$, these coefficients can be used to find the coefficients of the boundary Eq. (5). The values of $V^*$ used in (**c**) and (**d**) are the same as those used in (**b**) to plot $N$ and $K$, demonstrating that exponentially spaced $V^*$ give rise to uniformly spaced values in $(K,N)$.

**Hodgkin-Huxley boundary equation**

The coefficients of the HH boundary equation, Eq. (2), differ by a constant from those of the AK boundary, Eq. (5). The HH model is less excitable than the AK model, since less outward conductance is needed to suppress firing. In the approximations made in the 2D reduction of HH to AK, the largest source of error probably stems from the approximation of infinitely fast dynamics for the sodium activation variable, such that $m \approx m_\infty$. Around the typical voltage threshold, $m$ has a time constant of

~0.3-0.5 msec. This will cause the true *m* to lag behind the approximated $m \approx m_\infty$ and to take a lower value at any point in time leading up to the threshold, causing the increased excitability of the AK model. Comparing *m* dynamics during spiking near the boundary with those of $m_\infty$ (not shown), we find that multiplying $m_\infty$ by a constant factor of 3/4 provides a better approximation to *m* around the threshold. When this factor is included in the AK model, the boundary equation coefficients extracted from simulations are found to be 2.09 and 22.14, in good agreement with those determined by the HH equations in Eq. (2). With this correction, the *V*-nullcline of Eq. (9) can be used to analytically find the coefficients as above[5]: 2.07 and 21.3.

**Slow sodium inactivation**

While we have considered only steady state conductances, this approach may help to conceptualize the behavior of a system subject to certain types of adaptation, such as slow sodium inactivation (Arsiero et al., 2007). That the addition of slow sodium inactivation effectively lowers maximal sodium conductance for steady state can be seen from the model equations. Slow sodium inactivation may be introduced (Fleidervish et al., 1996; Miles et al., 2005) through an additional slow sodium inactivation variable *s*:

$$I_{Na} = G_{Na}m^3hs(V - E_{Na}), \qquad (14)$$

where *s* has the same form as the inactivation variable *h* but with a much slower time constant. The addition of *s* lowers the effective sodium pool for spiking. At steady state, when *s* is approximately constant due to its slow kinetics relative to spiking, this addition is equivalent to lowering $G_{Na}$. Through the sigmoidal voltage dependence of *s*, changes in mean current leads to effective changes of $G_{Na}$; as mean current increases, *s* decreases, decreasing the effective sodium pool. In other words, in contrast to spike-dependent adaptation currents, the voltage dependence of sodium inactivation provides feedback regardless of spiking activity. Thus, slow sodium inactivation, which in some sense is similar to voltage-dependent adaptation currents (Benda and Herz, 2003), is one mechanism by which the functional role of a neuron could change from integrating to differentiating without changes in channel density.

**Discussion**

Models of a variety of neurons share the same form as the Hodgkin-Huxley (HH) neuron (Ermentrout, 1998; Shriki et al., 2003), and evidence suggests that single-compartment models can capture key properties of *in vivo* and *in vitro* neurons (Destexhe et al., 2001). Here, we show specifically how a ratio of inward and outward conductances affect the HH neuron's sensitivity to input variance. The balance of outward to inward currents affects neuronal excitability. The well-known coincidence detection of auditory neurons (Reyes et al., 1996; Trussell, 1997) is enhanced by both a low-threshold potassium channel, which effectively decreases the membrane time constant (Reyes et al., 1994; Rathouz and Trussell, 1998; Rothman and Manis, 2003), and low availability of sodium channels (Svirskis et al., 2004). In dendrites, membrane excitability is modulated by changes in sodium or potassium currents (Colbert et al., 1997; Johnston et al., 1999). Low sodium conductance, which leads to spike frequency adaptation, may be the result of intrinsic low channel density (Melnick et al., 2004) or slow sodium inactivation (Fleidervish et al., 1996; Miles et al., 2005), which can be modulated by second messenger systems (Cantrell and Catterall, 2001).

---

[5] This is, of course, not a true *V*-nullcline of the HH model since we are not specifying a dependent variable to plot it against. But, as in the AK case, we assume that that the slow gating variables *n* and *h* are not functions of *V* and can then be ignored when partial differentiating with respect to *V*. This is analogous to assuming some sort of mapping from the 3D HH (i.e. excluding *m*) to a 2D model, and again we see that coefficients of Eq. (9), with the 3/4 correction factor multiplying $m_\infty$, are related to the coefficients of Eq. (2)

Neurons demonstrate a spectrum of integration and differentiation, and slow adaptation currents may correlate with shifts along this spectrum, at least for more complicated neurons such as neocortical pyramidal neurons (Higgs et al., 2006). However, diverging *f-I* curves, as in Figure 2 or Arsiero *et al.* (2007), would more likely be related to voltage-dependent rather than spike-dependent adaptation. Previous work has demonstrated that increased shunting, a likely consequence of strong excitatory and inhibitory synaptic input, can allow *M*-current adaptation to effectively switch a neuron's operating regime from integration to coincidence detection, or differentiation (Prescott et al., 2006). This may be a specific case of our general result, where the *M*-current essentially increases $G_K$ and shunting increases $G_{Leak}$. In some cases, the synaptic conductance alone may be large enough to switch a neuron from integration to differentiation. For example, Destexhe and Paré (1999) found that the input resistance of pyramidal cells in the parietal cortex decreases approximately five-fold during intense network activity. In the standard HH model, our results indicate that this change would place the cell near the boundary between integrator and differentiator function. Because synaptic conductances are highly variable on both short and long time scales, a postsynaptic neuron may fluctuate dynamically across the functional boundary.

Although we divide neural behavior into integration and differentiation as classified by the variance dependence of *f-I* curves, there are other possible ways to determine the computation that a single neuron performs on its current inputs. One example is white noise analysis, which we have explored extensively in applications to the Hodgkin-Huxley model (Aguera y Arcas et al., 2003), reduced models (Aguera y Arcas and Fairhall, 2003; Hong et al., In press), and neurons of avian brainstem (Slee et al., 2005). While we have found, suggestively, that the computation as described by a linear/nonlinear model changes as a function of $G_{Na}/G_K$ (R. Mease, unpublished data), here we do not attempt to connect these results.

In conclusion, we find that a plane separates two computational regimes in the space of maximal ionic conductances in the Hodgkin-Huxley and Abbott-Kepler models. Hyperplanes have been noted to separate models with different firing characteristics (Goldman et al., 2001; Taylor et al., 2006); in this case we show how the plane can be derived from characteristics of the *V* nullcline. Since *V* nullclines for reduced models of neuronal dynamics can in principle be obtained experimentally from *I(V)* plots (Izhikevich, 2007), this method may provide a means of relating biophysical properties with computational ones. Specifically, one may be able to infer certain biophysical parameters and their related effect on the neuron's computation from an experimentally obtained set of *I(V)* plots. Finally, although we focus here on maximal conductances, as might be regulated by homeostatic mechanisms, many forms of adaptation, such as slow sodium inactivation, change conductances over time, so that each adaptation state has a different maximal conductance. Time-dependent movement of the system through different computational states may be an effective way to describe the functional role of adaptation.

**Acknowledgments**

We thank W. J. Spain, Rebecca Mease, Michele Giugliano, and Larry Sorenson for helpful discussions and Michele Giugliano for comments on the manuscript. This work was supported by a Burroughs-Wellcome Careers at the Scientific Interface grant and a Sloan Research Fellowship to ALF; BNL was supported by the Medical Scientist Training Program, a fellowship from the National Institute of General Medical Sciences (T32 07266) and an ARCS fellowship; MHH was supported by a VA Merit Review to WJS.

## Appendix A: Biophysical Modeling

The single-compartmental conductance-based Hodgkin-Huxley (HH) model neuron (Hodgkin and Huxley, 1952) was used with standard parameters (Koch, 1999; Dayan and Abbott, 2001; Gerstner and Kistler, 2002) except as noted. In addition to Eq. (1), the following equations comprise the HH model:

$$\frac{dn}{dt} = \alpha_n(1-n) - \beta_n n,$$

$$\frac{dm}{dt} = \alpha_m(1-m) - \beta_m m,$$

$$\frac{dh}{dt} = \alpha_h(1-h) - \beta_h h,$$

$$\alpha_n(V) = \frac{0.01(V+55)}{1-e^{-0.1(V+55)}} \qquad \beta_n(V) = 0.125 e^{-(V+65)/80},$$

$$\alpha_m(V) = \frac{0.1(V+40)}{1-e^{-0.1(V+40)}} \qquad \beta_m(V) = 4 e^{-(V+65)/18},$$

$$\alpha_h(V) = 0.07 e^{-(V+65)/20} \qquad \beta_h(V) = \frac{1}{1+e^{-0.1(V+35)}},$$

where steady state gating values, such as $n_\infty$, are equal to $\alpha/(\alpha+\beta)$. Standard parameters for HH are: $G_{Na} = 120$, $G_K = 36$, and $G_{Leak} = 0.3$ mS/cm$^2$; $E_{Na} = 50$, $E_K = -77$, and $E_{Leak} = -54.4$ mV; and $C = 1$ µF/cm$^2$. For the cortical neuron model of Figure 1, standard parameters are: $G_{Na} = 50$, $G_{Kfast} = 225$, $G_{Kslow} = 0.225$, and $G_{Leak} = 0.25$ mS/cm$^2$; $E_{Na} = 74$, $E_{Kfast} = E_{Kslow} = -90$, and $E_{Leak} = -70$ mV; and $C = 1$ µF/cm$^2$.

Equations for the Abbott-Kepler model are derived from the HH model (Abbott and Kepler, 1990; Kepler et al., 1992; Hong et al., In press). Equations were solved numerically using fourth-order Runge-Kutta integration with a fixed time step of 0.05 msec, or 0.025 msec for the AK model. Injected current was simulated by a series of normally-distributed random numbers that were smoothed by an exponential filter ($\tau = 1$ msec). Spike times were identified as the upward crossing of the voltage trace at -20 mV (resting potential = -65 mV) separated by more than 2 msec.

## Appendix B: Examining $f(U,V)$

That Eq. (9) is the appropriate condition can also be seen as follows. While the function $f$ depends on both $U$ and $V$, along the nullclines, $U$ is a function of $V$, $U=g(V)$. Thus, requiring $df(U,V)/dV = 0$ at the fixed point as required by the definition of a $V$ nullcline:

$$\frac{df(U,V)}{dV} = \frac{\partial f}{\partial V} + \frac{\partial f}{\partial U}\frac{dU}{dV} = 0,$$

$$\frac{dU}{dV} = -\frac{\frac{\partial f}{\partial V}}{\frac{\partial f}{\partial U}} = 0, \qquad (15)$$

where $\partial f/\partial U > 0$. Then, the numerator must equal zero, giving Eq. (9). One can show that the full stability analysis around a fixed point gives almost the same result as our discussion based on the shape of a nullcline. For the 2D AK model, the following two conditions on the determinant and trace of the Jacobian matrix $J$ evaluated at the fixed point determine the stability of the fixed point (Strogatz, 1994):

$$\det(J) = \frac{\partial F}{\partial V}\frac{\partial g}{\partial U} - \frac{\partial F}{\partial U}\frac{\partial g}{\partial V} > 0,$$

$$\mathrm{Tr}(J) = \frac{\partial F}{\partial V} + \frac{\partial g}{\partial U} < 0,$$

(16)

where $F(U,V) = -f(U,V)/C + I/C = dV/dt$, given Eq. (7), and $g(U,V) = dU/dt$. The first condition is always satisfied for AK models, as the Jacobian at fixed points of these models always have complex eigenvalues due to their resonating dynamics. For the $U$ nullcline of the AK model, $g \approx -k(U-V)$, where $k$ has the form:

$$k = \frac{G_K F_K(V) t_K(V) + G_{Na} F_{Na}(V) t_{Na}(V)}{G_K F_K(V) + G_{Na} F_{Na}(V)}.$$

The functions $t_i(V)$ are slowly varying functions that are of the same order of magnitude over the range of $V$ that we are interested in. Therefore, $k$ weakly depends on conductances and is roughly constant in $V$. Moreover, the value of $k$ is small with a maximum value less than 0.3 mV/msec. Therefore, $\frac{\partial f}{\partial V} = -C\frac{\partial F}{\partial V} \lesssim 0.3$ mS/cm$^2$ is approximately equivalent to a full stability analysis based on the Jacobian. Even when $k$ is nonzero, there isn't a qualitative change; since $\frac{\partial f}{\partial V} = G_{Leak} + (V \text{ dependent terms})$, considering non-zero constant $k$ is equivalent to $G_{Leak} \to G_{Leak} + k$, which corresponds to a small offset of our boundary plane such that it does not pass through the origin. Thus, the right hand side of Eqs. (2) and (5) may not equal zero. When we fit our simulated data and allow for a nonzero offset, the coefficients and error of the fit are almost identical as previously, and the fitted offset is small (HH: ~-0.3; AK: ~1.0).